\documentclass{article} % For LaTeX2e
\usepackage{iclr2026_conference,times}
\usepackage{amsmath,amsfonts,bm}

\def\eqref#1{equation~\ref{#1}}
\def\1{\bm{1}}

\DeclareMathAlphabet{\mathsfit}{\encodingdefault}{\sfdefault}{m}{sl}
\SetMathAlphabet{\mathsfit}{bold}{\encodingdefault}{\sfdefault}{bx}{n}

\usepackage[bookmarks=true]{hyperref}
\usepackage{graphicx}
\usepackage{url}

\usepackage[utf8]{inputenc} % allow utf-8 input
\usepackage[T1]{fontenc}    % use 8-bit T1 fonts
\usepackage{hyperref}       % hyperlinks
\usepackage{url}            % simple URL typesetting
\usepackage{booktabs}       % professional-quality tables
\usepackage{amsfonts}       % blackboard math symbols
\usepackage{nicefrac}       % compact symbols for 1/2, etc.
\usepackage{microtype}      % microtypography
\usepackage{color}          % colors

\usepackage{xcolor}
\usepackage{cancel}
\usepackage{amsmath,amssymb}
\allowdisplaybreaks[3] % 1(mild) - 4(positive)
\usepackage{amsthm}
\usepackage{ascmac}
\usepackage{graphicx}
\usepackage{bm}
\usepackage{comment}
\usepackage{environ}
\usepackage{fancybox}
\usepackage{setspace}
\usepackage{verbatim}
\usepackage{algorithm} % for algorithm environment
\usepackage{algpseudocode} % for algorithmic environment
\definecolor{myblue}{rgb}{0.2, 0.2, 0.7}
\definecolor{migraine}{rgb}{0.0, 0.5, 0.0}

\newcommand{\sgn}{\,\mbox{\rm sgn}\,}
\newcommand{\erf}{\,\mbox{\rm erf}\,}

\newcommand{\rmi}{\mathrm{i}}

\newcommand{\llangle}{\langle\kern -.23em \langle}
\newcommand{\rrangle}{\rangle\kern -.23em \rangle}

\renewcommand{\vec}[1]{\boldsymbol{#1}}

\newtheorem{definition}{Definition}
\newtheorem{proposition}{Proposition} %\proof
\newtheorem{assumption}{Assumption}
\newtheorem{lemma}{Lemma}
\newtheorem{corollary}{Corollary}

\title{Dynamical Properties of Dense Associative Memory}
\author{%
	Kazushi Mimura \\
	Hiroshima City University, Hiroshima, Japan \\
	RIKEN AIP, Tokyo, Japan \\
	\texttt{\small mimura@hiroshima-cu.ac.jp} \\
	\texttt{\small kazushi.mimura@a.riken.jp} \\
	\And
	Jun'ichi Takeuchi \\
	Graduate School of Info. Sci. and Elect. Eng. \\
	Kyushu University, Fukuoka, Japan \\
	\texttt{\small tak@inf.kyushu-u.ac.jp} \\
	\AND
	Yuto Sumikawa \\
	Institute for Physics of Intelligence \\
	The University of Tokyo, Tokyo, Japan \\
	\texttt{\small sumikawa-yuto@g.ecc.u-tokyo.ac.jp}\\
	\And
	Yoshiyuki Kabashima \\
	Institute for Physics of Intelligence \\
	The University of Tokyo, Tokyo, Japan \\
	\texttt{\small kaba@phys.s.u-tokyo.ac.jp} \\
	\And
	Anthony C. C. Coolen \\
	Saddle Point Science Europe and DCN Donders Institute,  \\
	Radboud University, Nijmegen, Netherlands \\
	\texttt{\small a.coolen@science.ru.nl} \\
}
\iclrfinalcopy % Uncomment for camera-ready version, but NOT for submission.

\begin{document}
\maketitle
% storage capacity
\begin{abstract}
Dense associative memory, a fundamental instance of modern Hopfield networks, can store a large number of memory patterns as equilibrium states of recurrent networks. While the stationary-state storage capacity has been investigated, its dynamical properties have not yet been discussed. In this paper, we analyze the dynamics using an exact approach based on generating functional analysis. We show results on convergence properties of memory retrieval, such as the convergence time and the size of the attraction basins. Our analysis enables a quantitative evaluation of the convergence time and the storage capacity of dense associative memory, which is useful for model design. Unlike the traditional Hopfield model, the retrieval of a pattern does not act as additional noise to itself, suggesting that the structure of modern networks makes recall more robust. Furthermore, the methodology addressed here can be applied to other energy-based models, and thus has the potential to contribute to the design of future architectures.
\end{abstract}

%--------------------------------------------------
\section{Introduction}
\par

%--------------------------------------------------
\subsection{Background}
\par
Dense associative memory \citep{Krotov2016}, a model for storing binary patterns, was proposed and shown to significantly improve the storage capacity of the traditional Hopfield model \citep{Hopfield1982}. While it can be regarded as a rediscovery of the many-body Hopfield model \citep{Gardner1987, Abbott1987}, it exhibits slightly different properties. On the other hand, extensions of dense associative memory, such as the Hopfield layer, have been actively developed to enable dense associative memory to store real-valued patterns \citep{Demircigil2017, Ramsauer2021}. Hopfield models with such large memory capacities, including these variants, are referred to as modern Hopfield networks, which have gained increasing attention and have even inspired Transformer architectures \citep{Hoover2023}. 
\par
The equilibrium properties of the Hopfield layer have been analytically studied, including evaluations of its storage capacity \citep{Lucibello2024}. Since the Hopfield layer can reach a near-equilibrium state in almost a single update step, its dynamical properties have not been considered a significant issue. In contrast, dense associative memory, like the traditional Hopfield model, requires some iterative updates to reach a stationary state. However, its dynamical behavior has not been investigated so far. As a result, fundamental aspects such as the attraction basin, namely, how far from a stored pattern can the initial state be for the system to still successfully recall it, still remain unclear. 
\par
While the dynamical properties of dense associative memory have not been investigated, those of the traditional Hopfield model have been extensively analyzed. In this paper, we analyze the dynamical properties of dense associative memory using generating functional analysis, an asymptotic theory in the large-system limit, which has been widely used in those studies.
%--------------------------------------------------
\subsection{Contributions}
\par
Our main contributions are as follows:
\begin{itemize}
	\item 
	Asymptotically exact dynamical analysis. -- 
	We provide, for the first time, an asymptotically exact analysis 
	of the dynamics of dense associative memory in the large-system limit 
	using generating functional analysis (GFA).
	\item 
	Quantitative characterization of convergence. --  
	Our analysis yields explicit results on convergence properties of memory retrieval, 
	including convergence time and the size of attraction basins, 
	thereby enabling quantitative evaluation of stability and storage capacity. 
	\item 
	Novel insight into robustness of modern Hopfield networks. -- 
	We demonstrate that, unlike the traditional Hopfield model, 
	retrieval does not introduce additional self-noise, 
	suggesting that the architecture of modern networks makes recall more robust. 
	\item 
	General methodology for energy-based models. -- 
	The proposed framework is not limited to dense associative memory. 
	It can be applied to other energy-based models, 
	providing theoretical tools for the design of robust and scalable architectures.
\end{itemize}
\subsection{Related Works}
\par
Gardner and Abbott independently introduced a Hopfield model with many-body interactions, which is essentially equivalent to dense associative memory, and analyzed its equilibrium properties using the replica method to evaluate its storage capacity. The difference between their models and the dense associative memory proposed by Krotov and Hopfield lies in the presence or absence of self-coupling terms. While this difference does not affect the order of the storage capacity, it does influence the constant factor. Additionally, Lucibello and M\'ezard analyzed the equilibrium properties of the Hopfield layer using the replica method and obtained its storage capacity \citep{Lucibello2024}. So far, no analysis of the dynamical behavior of the modern Hopfield model has been reported. On the other hand, there has been extensive research on the dynamics of the traditional Hopfield models. For example, the papers \citep{Gardner1987-2, Crisanti1987, Crisanti1988, Rieger1988, During1998, Coolen2000, Mimura2004, Mimura2009} provide exact analysis based on GFA.

%--------------------------------------------------
\section{Preliminalies}
\par

%--------------------------------------------------
\subsection{Notations}
\par
Throughout this paper, vectors are denoted by boldface, e.g., $\vec{x}$, and are assumed to be column vectors unless otherwise stated. $x^{(t)}$ and $(\vec{x})^{(t)}$ represent the $t$-th element of the vector $\vec{x}$. Matrices are denoted by an upper case symbol, e.g., $A$, and $A^\top$ denotes the transpose of a matrix $A$.

%--------------------------------------------------
\subsection{Dense Associative Memory}
\par
The dense associative memory is one of the recurrent neural network models to store and recall a large number of patterns as fixed points of the dynamics. The energy of dense associative memory is given by 
\begin{equation}
	H = -\sum_{\mu=1}^M F \biggl( \sum_{i=1}^N \xi_i^\mu h_i \biggr), 
\end{equation}
where $h_i \in \{\pm1\}$ denotes the state of the $i$-th unit, and $\xi_i^\mu$ denotes the $i$-th element of the $\mu$-th pattern. Each $\xi_i^\mu$ independently takes the value $\pm1$ with equal probability $1/2$. Introducing a nonlinear function $F$, such as a power function, makes memory patterns become deeper minima in the energy landscape and reduces interference between different memory patterns. This is because the nonlinearity suppresses weaker overlaps during the recall process. In this paper, we restrict ourselves to the case 
\begin{equation}
	F(x)=\frac{x^n}{2 N^{n-1}}. 
\end{equation}
It should be noted that since the coefficient $1/(2 N^{n-1})$ does not affect the performance, so it is equivalent to setting $F(x)=x^n$. The update rule is defined by the difference of two energies. before and after state transitions. We keep only the leading term in the argument of the sgn function in the update rule, which gives 
\begin{align}
	h_i^{(t+1)} 
	=& 
	\sgn \biggl[
	\sum_{\mu=1}^M F \biggl( + \xi_i^\mu + \sum_{j \ne i}^N \xi_j^\mu h_j \biggr) - 
	\sum_{\mu=1}^M F \biggl( - \xi_i^\mu + \sum_{j \ne i}^N \xi_j^\mu h_j \biggr) 
	\biggr]
	\\
	=& 
	\sgn \biggl[
	\sum_{\mu=1}^M 
	\xi_i^\mu n \biggl( \frac1N \sum_{j \ne i}^N \xi_j^\mu h_j \biggr)^{n-1}
	+ \mbox{(small order terms)} \;
	\biggr], 
\end{align}
where $\sgn(x)$ denotes the sign function that takes $1$ if $x \ge 0$, and $-1$ otherwise. In the case of $n=2$ the network reduces to the parallel dynamics version of traditional Hopfield model, i.e., $h_i^{(t+1)} = \sgn ( \sum_{j=1}^N J_{ij} h_j^{(t)} )$ and $J_{ij} = \frac1{N} \sum_{\mu=1}^M \xi_i^\mu \xi_j^\mu$.

%--------------------------------------------------
\subsection{Outline of Generating Functional Analysis}
\par
We apply the generating functional analysis (GFA) to investigate dynamical properties of the dense associative memory. GFA has been applied to the model which is described using realizations of random variables \citep{DeDominicis1978}. This method allows us to analyze the asymptotic dynamical behavior in the infinitely large system, using the generating functional, which is the dynamical equivalent of the characteristic function in statistics. 
\par
In GFA formalism, we consider the joint probability distribution over the states of all units at all time steps, from the start of the iteration up to some prescribed time, which can be taken sufficiently large. This joint probability is referred to as the \textit{path probability}. From the path probability, we can calculate various expectation values such as the \textit{overlap}, which is the direction cosine between the states of the units and the memory pattern being recalled via the \textit{generating functional} which can be regarded as an analogue of the characteristic function.

%--------------------------------------------------
\section{Analysis}
\par
First, the path probability is defined and used to describe the generating functional, after which the expectation over the memory patterns appearing in the generating functional is evaluated. 
%In GFA, we consider the joint probability  the states of all units from initial condition a given finite time$p[\vec{h}^{(0)}, \cdots, \vec{h}^{(T)}]$, we can use it to calculate various expectation values. This joint probability is referred to as the {\it path probability}. The way to evaluate expected values of interest, we introduce the \textit{generating functional}, defined as $Z[\vec{\psi}]=$ $\langle$ $\exp [ - \rmi \sum_{t=0}^{T} \sum_{i=1}^N \psi_i^{(t)} h_i^{(t)}] \rangle$, where the bracket $\langle \cdots \rangle$ denotes the average over the path probability $p[\vec{h}^{(0)}, \cdots, \vec{h}^{(T)}]$, and we have introduced the generating variables $\vec{\psi}^{(t)}=(\psi_i^{(t)})$ $\in \mathbb{R}^N$ and we write $\vec{\psi}=(\vec{\psi}^{(0)}, \cdots, \vec{\psi}^{(T)})$ for shorthand. 

%--------------------------------------------------
\subsection{Path Probability}
\par
Let vectors $\vec{h}^{(t)}=(h_1^{(t)}, \cdots, h_N^{(t)})^\top$ $\in \{\pm 1\}^N$ be the states of all units at time $t$ and let the initial state be $\vec{h}^{(0)}$. The updating rule, obtained by retaining only the leading term, is expressed as follows: 
\begin{align}
	& h_i^{(t+1)} = \sgn(u_i^{(t)}), \\
	& u_i^{(t)} = \sum_{\mu=1}^M 
	\xi_i^\mu n \biggl( \frac1N \sum_{j \ne i}^N \xi_j^\mu h_j^{(t)} + \theta_i^{(t)} \biggr)^{n-1}, 
\end{align}
% where $\vec{u}^{(t)}=(u_1^{(t)}, \cdots, u_N^{(t)})^\top$ and $\sgn$ is entrywise applied for a vector. 
for all $i \in \{1, \cdots, N \}$ and $t \in \{0,\cdots,T-1\}$. The variable $u_i^{(t)}$ is referred to as a {\it local field}. The parameter $\theta_i^{(t)}$ is called an \textit{external field} (or an \textit{threshold}). The dynamics of the system are characterized by how the output of each unit changes in response to infinitesimal variations in the local field. To evaluate such changes, the external field $\{\theta_i^{(t)}\}$ are introduced. The average derivative of the outputs with respect to the external field is referred to as the response function, which serves as one of fundamental measures for describing the dynamics. After evaluating the response function, all $\{\theta_i^{(t)}\}$ are set to zero. 
\par
In this paper, we consider parallel dynamics, in which the states of all units are updated simultaneously. The updating rule of the dense associative memory for the variable $\vec{h}^{(t+1)}$ at time $t$ can be given by the following probability distribution: 
\begin{align}
	p[\vec{h}^{(t+1)}|\vec{h}^{(t)}] = \prod_{i=1}^N \delta [ h_i^{(t+1)}; \sgn(u_i^{(t)}) ], 
\end{align} 
where $\delta[m;n]$ denotes the Kronecker's delta that takes $1$ if $m=n$ and $0$ otherwise. This dynamics represents Markovian dynamics. The path probability $p[\vec{h}^{(0)},\cdots,\vec{h}^{(T)}]$ is given as the products of the probability distribution of the updating rule: 
\begin{align}
	p[\vec{h}^{(0)},\cdots,\vec{h}^{(T)}] 
	= 
	p[\vec{h}^{(0)}] \prod_{t=0}^{T-1} p[\vec{h}^{(t+1)}|\vec{h}^{(t)}], 
	\label{eq:def_path_probability}
\end{align}
where $p[\vec{h}^{(0)}]=\prod_{i=1}^N p[h_i^{(0)}]$ denotes the initial state distribution. Since the same memory patterns are included at every time step, the states of the units at different times are correlated.

%--------------------------------------------------
\subsection{Generating Functional}
\par
The path probability depends on all memory patterns $\vec{\xi}^1, \cdots, \vec{\xi}^M$. We define the generating functional as follows. 
\begin{definition} % DEFINITION
	\label{definition:Z}
	The generating functional $\bar{Z}[\vec{\psi}]$ is defined as 
	\begin{align}
		\bar{Z}[\vec{\psi}]
		= 
		\mathbb{E}_{\vec{\xi}^1, \cdots, \vec{\xi}^M}
		\biggl[
		\sum_{\vec{h}^{(0)}, \cdots, \vec{h}^{(T)} \in \{\pm 1\}^N}
		p[\vec{h}^{(0)},\cdots,\vec{h}^{(T)}] 
		\exp \biggl(-i\sum_{t=0}^T\vec{h}^{(t)}\cdot\vec{\psi}^{(t)} \biggr) 
		\biggr], 
		\label{eq:def-Z}
	\end{align}
	where we have introduced the generating variables 
	$\vec{\psi}^{(t)}=(\psi_1^{(t)}, \cdots, \phi_N^{(t)})^\top$ $\in \mathbb{R}^N$ 
	and we write $\vec{\psi}=(\vec{\psi}^{(0)}, \cdots, \vec{\psi}^{(T)})$ for shorthand. 
\end{definition}
Here, $i$ denotes the imaginary unit. i.e., $i=\sqrt{-1}$. We here assumed that the generating functional is self-averaging, namely, in the large-system limit, i.e., $N$ is sufficiently large, the generating functional is concentrated on its average over the memory patterns $\vec{\xi}^1, \cdots, \vec{\xi}^M$ and the typical behaviour of the model only depends on the statistical properties of the memory patterns. In GFA, the expectation values of interest are calculated from derivatives with respect to some elements of the generating variables, e.g., 
\begin{align}
	& 
	\lim_{\vec{\psi}\to\vec{0}} 
	\frac{\partial \bar{Z}[\vec{\psi}]}{\partial \psi_i^{(t)}} 
	= 
	\mathbb{E}_{\vec{\xi}^1, \cdots, \vec{\xi}^M} 
	[\langle -i h_i^{(t)} \rangle], 
	\label{eq:derivatives.1}
	\\
	&
	\lim_{\vec{\psi}\to\vec{0}} 
	\frac{\partial^2 \bar{Z}[\vec{\psi}]}{\partial \psi_i^{(t)} \partial \psi_i^{(t')}} 
	= 
	\mathbb{E}_{\vec{\xi}^1, \cdots, \vec{\xi}^M} 
	[\langle - h_i^{(t)} h_{i'}^{(t')} \rangle], \\
	&
	\lim_{\vec{\psi}\to\vec{0}} 
	\frac{\partial^2 \bar{Z}[\vec{\psi}]}{\partial \psi_i^{(t)} \partial \theta_i^{(t')}} 
	= 
	\mathbb{E}_{\vec{\xi}^1, \cdots, \vec{\xi}^M} [\langle 
	- i 
	\frac{\partial h_i^{(t)} }{\partial \theta_{i'}^{(t')}} \rangle], 
	\label{eq:derivatives.3}
\end{align}
where $\vec{\psi}\to\vec{0}$ denotes $\psi_i^{(t)} \to 0$ for all $i$ and $t$, and the bracket $\langle \cdots \rangle$ denotes the average over the path probability, i.e., $\langle (\cdots) \rangle =$ $\sum_{\vec{h}^{(0)}, \cdots, \vec{h}^{(T)} \in \{\pm 1\}^N}$ $p[\vec{h}^{(0)},\cdots,\vec{h}^{(T)}] (\cdots)$. Introducing the definition of the local field using the Dirac delta function, the generating functional can be rewritten as follows: 
\begin{align}
	\bar{Z}[\vec{\psi}]
	=& 
	\mathbb{E}_{\vec{\xi}^1, \cdots, \vec{\xi}^M}
	\biggl[ 
	\sum_{\vec{h}^{(0)}, \cdots, \vec{h}^{(T)}}
	\int_{\mathbb{R}^T} d\vec{u}~
	p[\bm{h}^{(0)}]
	\biggl( 
		\prod_{t=0}^{T-1} \prod_{i=1}^N 
		\delta [h_i^{(t+1)} ; \mathrm{sgn}(u_i^{(t)}) ] 
	\biggr) 
	e^{ - i \sum_{t=0}^T \bm{h}^{(t)} \cdot \bm{\psi}^{(t)} }
	\nonumber\\
	& 
	\times \biggl( 
		\prod_{t=0}^{T-1} 
		\prod_{i=1}^N 
		\delta \biggl( u_i^{(t)} 
		- \sum_{\mu =1}^M \xi_i^\mu 
		\biggl( \frac1N \sum_{j \ne i}^N \xi_j^\mu h_j^{(t)} \biggr)^{n-1}
		- \theta_i^{(t)}
		\biggr)
	\biggr)
	\biggr]. 
\end{align}
\par
Assuming that the pattern $\vec{\xi}^1$ is being recalled, we separate the local field in the generating functional into a signal term including the recalling pattern and a noise term including other patterns $\vec{\xi}^2, \cdots, \vec{\xi}^M$. Using the Fourier integral form of Dirac delta function, the generating functional becomes
\begin{align}
	\bar{Z}[\vec{\psi}]
	=& 
	\sum_{\vec{h}^{(0)}, \cdots, \vec{h}^{(T)}}
	\int_{\mathbb{R}^T} d\vec{u} \delta\hat{\vec{u}}~
	p[\vec{h}^{(0)}]
	\biggl( 
		\prod_{t=0}^{T-1} \prod_{i=1}^N 
		\delta [h_i^{(t+1)} ; \mathrm{sgn}(u_i^{(t)}) ] 
	\biggr) 
	e^{ - i \sum_{t=0}^T \bm{h}^{(t)} \cdot \bm{\psi}^{(t)} }
	\nonumber\\
	& 
	\times \exp \biggl[ i \sum_{t=0}^T \sum_{i=1}^N \hat{u}_i^{(t)} (u_i^{(t)} - \theta_i^{(t)})\biggr] 
	\biggl( \mathbb{E}_{\vec{\xi}^1}
	\exp \biggl[ - i \sum_{t=0}^{T-1} \sum_{i=1}^N 
		\hat{u}_i^{(t)} \xi_i^1 ~ 
		n \biggl( \frac1N \sum_{j \ne i}^N \xi_j^1 h_j^{(t)} \biggr)^{n-1}
	\biggr] \biggr) \notag\\
	&
	\times 
	\biggl( \mathbb{E}_{\vec{\xi}^2, \cdots, \vec{\xi}^M}	
	\exp \biggl[ - i \sum_{t=0}^{T-1} \sum_{i=1}^N 
		\sum_{\mu =2}^M 
		\hat{u}_i^{(t)} \xi_i^\mu ~ 
		n \biggl( \frac1N \sum_{j \ne i}^N \xi_j^\mu h_j^{(t)} \biggr)^{n-1}
	\biggr] \biggr)
	\label{eq:noise-in-Z}
\end{align}
Only the last part involves all non-recalled patterns $\vec{\xi}^2, \cdots, \vec{\xi}^M$. By straightforward calculation of the expectation over these patterns, the generating functional is found to depend on five types of averages. Accordingly, we introduce the following macroscopic parameters: 
\begin{align}
	& 
	m^{(t)} = \frac1N \sum_{i=1}^N \xi_i h_i^{(t)}, ~~ 
	k^{(t)} = \frac1N \sum_{i=1}^N \xi_i \hat{u}_i^{(t)}, \notag\\
	&q^{(t,t')} = \frac1N \sum_{i=1}^N h_i^{(t)} h_i^{(t')}, ~~ 
	Q^{(t,t')} = \frac1N \sum_{i=1}^N h_i^{(t)} \hat{u}_i^{(t')}, ~~ 
	K^{(t,t')} = \frac1N \sum_{i=1}^N h_i^{(t)} \hat{u}_i^{(t')}, 
	\label{eq:macroscopic-parameters}
\end{align}
into the generating functional using the Dirac delta function, where the functions $m^{(t)}$ is referred to as the \textit{overlap}. The generating functional can be calculated as follows. 

\begin{lemma}
	\label{lemma:Z}
	By averaging over the memory patterns, the generating functional is given by 
	\begin{align}
		\bar{Z}[\bm{\psi}]
		= 
		\int
		d\vec{m} d\hat{\vec{m}}
		d\vec{k} d\hat{\vec{k}}
		d\vec{q} d\hat{\vec{q}}
		d\vec{Q} d\hat{\vec{Q}}
		d\vec{K} d\hat{\vec{K}} ~ 
		%e^{N(\Psi+\Phi+\Omega)+O(\log N)}, 
		\exp \biggl[ N(\Psi+\Phi+\Omega) + O(\log N) \biggr], 
		\label{eq:bar-Z-in-lemma}
	\end{align}
	where 
	\begin{align}
		\Psi
		=& 
		%\textstyle
		i \sum_{t=0}^{T-1} \biggl\{ 
		\hat{m}^{(t)} m^{(t)} + 
		\hat{k}^{(t)} k^{(t)} - 
		k^{(t)} ~ n ( m^{(t)} )^{n-1}
		\biggr\} 
		\notag\\
		&
		%\textstyle
		+ i \sum_{t=0}^{T-1} \sum_{t'=0}^{T-1} \biggl\{ 
		\hat{q}^{(t,t')} q^{(t,t')} + 
		\hat{Q}^{(t,t')} Q^{(t,t')} + 
		\hat{K}^{(t,t')} K^{(t,t')} \biggr\}, \\
	%\end{align}
	%\begin{align}
		\Phi
		=& 
		%\textstyle
		\frac1N
		\log
		\sum_{\vec{h}} \int d\vec{u} d\hat{\vec{u}} ~ 
		p[\vec{h}^{(0)}]
		\biggl( 
			\prod_{t=0}^{T-1} \prod_{i=1}^N 
			\delta [h_i^{(t+1)} ; \mathrm{sgn}(u_i^{(t)}) ] 
		\biggr) 
		\nonumber\\
		& 
		%\textstyle
		\times E_{\vec{\xi}} \exp \biggl[
		-i \sum_{t=0}^{T-1} \sum_{t'=0}^{T-1} \sum_{i=1}^N \biggl\{
		\hat{q}^{(t,t')} h_i^{(t)} h_i^{(t')} + 
		\hat{Q}^{(t,t')} \hat{u}_i^{(t)} \hat{u}_i^{(t')} + 
		\hat{K}^{(t,t')} h_i^{(t)} \hat{u}_i^{(t')} \biggr\}
		\nonumber\\
		& 
		%\textstyle
		+ i \sum_{t=0}^{T-1} \sum_{i=1}^N \hat{u}_i^{(t)} 
		\{ u_i^{(t)} - \hat{k}^{(t)} - \theta_i^{(t)} \}
		- i \sum_{t=0}^{T-1} \sum_{i=1}^N h_i^{(t)} \hat{m}^{(t)} \xi_i
		- i \sum_{t=0}^{T-1} \sum_{i=1}^N h_i^{(t)} \psi_i^{(t)} 
		\biggr], \\
	%\end{align}
	%\begin{align}
		\Omega
		=& 
		%\textstyle
		-\frac12 n^2 \frac{M}{N^{n-1}} %\alpha_n
		\sum_{t=0}^{T-1} \sum_{t'=0}^{T-1} \biggl\{ 
		(n-1)^2 
		\biggl[ ~ \sum_{k=0}^{n-2} A(n-2,k) 
		( q^{(t,t')} )^k \biggr] 
		K^{(t',t)}
		K^{(t,t')}
		\nonumber\\
		& 
		%\textstyle
		+ 
		\biggl[ ~ \sum_{k=0}^{n-1} A(n-1,k) 
		( q^{(t,t')} )^k \biggr]
		Q^{(t',t)} 
		\biggr\} +O(N^{-1}). 
	\end{align}
	where 
	$	
		A(\ell,k)
		=
		\bigl(
		\!\!
		\begin{array}{c}
			\ell \\
			k \\
		\end{array}
		\!\!
		\bigr)^2 \; k! \; B(\ell-k)^2
	$, and 
	$
		B(m) 
		= 
		\vec{1}_{m\mbox{:even}} \; (m-1)!!.  
	$
\end{lemma}
A proof sketch is given in Appendix \ref{appendix:Z}. Here, $\vec{1}_{\mathrm{condition}}$ denotes the indicator function that takes $1$ if the condition is true, and $0$ otherwise. It can be obtained by evaluating the leading terms after taking the expectation over the memory patterns, using combinatorial arguments. The order of the number of memory patterns is determined by the balance between the magnitude of the signal originating from the retrieved pattern and that of the noise originating from the non-retrieved patterns. From the analysis in Lemma \ref{lemma:Z}, the number of memory patterns $M$ is required to scale as $M=O(N^{n-1})$ for non-trivial analysis. This corresponds to the generating functional being of order $e^{O(N)}$. Therefore, we set 
\begin{align}
	M = \alpha_n N^{n-1}. 
\end{align}
Further details are given in Appendix \ref{appendix:Z}. 
\par
The generating functional is dominated by a saddle-point in the large-system limit. Averaging over the random variables, we will move to a saddle-point problem \citep{Copson1965} in the limit $N \to \infty$. The saddle point condition gives values of the macroscopic parameters. Hereafter, we choose the factorised distribution $p[\vec{h}^{(0)}]$ $=$ $\prod_{i=1}^N$ $p[h_i^{(0)}]$ $=$ $\prod_{i=1}^N \{ \frac12(1+m^{(0)}) \delta[h^{(0)};\xi_i] + \frac12(1-m^{(0)}) \delta[ h^{(0)}; - \xi_i]\}$ as an initial state distribution, where $m^{(0)}$ denotes an \textit{initial overlap}. The factorised initial overlap allows the generating functional to decompose into independent single-unit contributions. 
%The integrand in the generating functional (\ref{eq:Z}) scales as $e^{O(N)}$. In order to obtain a nontrivial storage capacity, the order of $\mathcal{N}$ must be $e^{O(N)}$. From this, the scaling of $M$ can be determined analytically, which is equivalent to the discussion of the absolute capacity by Krotov and Hopfield. The main part of our analysis is the calculation of $\mathcal{N}$. 

%--------------------------------------------------
\section{Main results}
\par

%--------------------------------------------------
%\subsection{Exact dynamics}
%\par
The behavior of this model differs significantly between the case $n=2$ and the case $n \ge 3$. Since the case $n=2$ has already been extensively studied, we focus only on the case $n \ge 3$ in this paper. GFA provides an exact solution as an asymptotic analysis in the large-system limit $N \to \infty$. Using the saddle point method to evaluate the integral in the averaged generating functional, one can obtain the following proposition.

\begin{proposition} % PROPOSITION
	\label{proposition:gfa}
	For a given initial state distribution $p[h^{(0)}]$ and $n \ge 3$, 
	the overlap $m^{(t)}$, the correlation function $C^{(t,t')}$, 
	and the response function $G^{(t,t')}$ are given by 
	\begin{align}
		m^{(t)} = \llangle \xi h^{(t)} \rrangle, ~~~~ 
		C^{(t,t')} = \llangle h^{(t)} h^{(t')} \rrangle, ~~~~ 
		G^{(t,t')} = \vec{1}_{t>t'}\;\frac{\partial\llangle h^{(t)}\rrangle}{\partial\theta^{(t')}},
	\end{align}
	where $\llangle f(\vec{h}) \rrangle$ denotes the average defined as 
	\begin{align}
		\llangle f(\vec{h}) \rrangle 
		=
		\mathbb{E}_\xi \int \mathcal{D}\vec{v} \sum_{\vec{h}} f(\vec{h})
		p[h^{(0)}] \prod_{t=0}^{T-1}
		\delta \biggl[ 
			h^{(t+1)} ; 
			\sgn \biggl( \xi n(m^{(t)})^{n-1} 
			+ (\Gamma \vec{h})^{(t)} + v^{(t)} + \theta^{(t)} \biggr)
		\biggr], 
		\label{eq:effective-path-measure}
	\end{align}
	which is referred to as the \textit{effective path measure}. 
	The random vector $\vec{v}$ follows a multivariate normal distribution 
	with mean $\vec{0}$ and covariance matrix $R=(R^{(t,t')})$, where the $(t,t')$-element is 
	\begin{align}
		R^{(t,t')} =& n^2 \alpha_n \sum_{k=0}^{n-1} A(n-1,k) (C^{(t,t')})^k. 
	\end{align}
	The matrix $\Gamma$ is given by $\Gamma = D \circ G$. 
	The $(t,t')$-elements of $D$ and $G$ are $D^{(t,t')}$ and $G^{(t,t')}$, respectively. 
	Each element of the matrix $D=(D^{(t,t')})$ is defined as 
	\begin{align}
		D^{(t,t')} =& n^2 (n-1)^2 \alpha_n \sum_{k=0}^{n-2} A(n-2,k) (C^{(t,t')})^k. 
	\end{align}
	The operator $\circ$ denotes the Hadamard (elementwise) product. 
\end{proposition}
The proof sketch is given in Appendix \ref{appendix:gfa}. The term $(\Gamma \vec{h})^{(t)}$ in the effective path measure represents a \textit{retarded self-interaction}. Due to this retarded self-interaction, the state at the next time step depends in a complex way on the past states. On the other hand, unlike in the traditional Hopfield model, i.e., the case of $n=2$, the noise variance does not depend on the overlap, and it does not increase even when the overlap becomes large. As a result, it is considered that the phenomenon, in which the system begins to recall correctly but eventually fails to complete it, becomes less likely to occur.

\begin{figure}[t]
	\centering
	\includegraphics[width=0.45\linewidth,keepaspectratio]
	{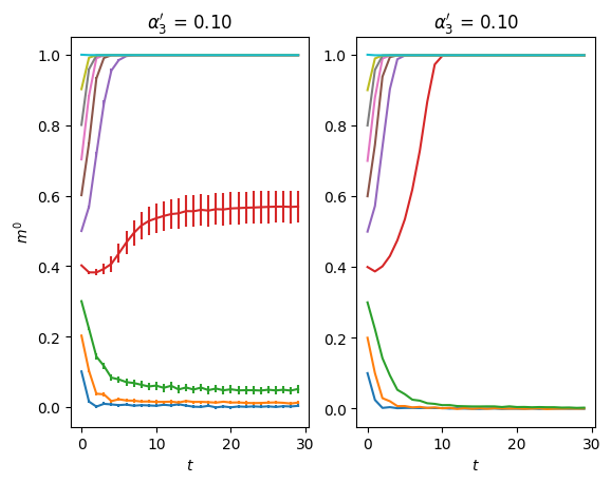} ~~~~
	\includegraphics[width=0.45\linewidth,keepaspectratio]
	{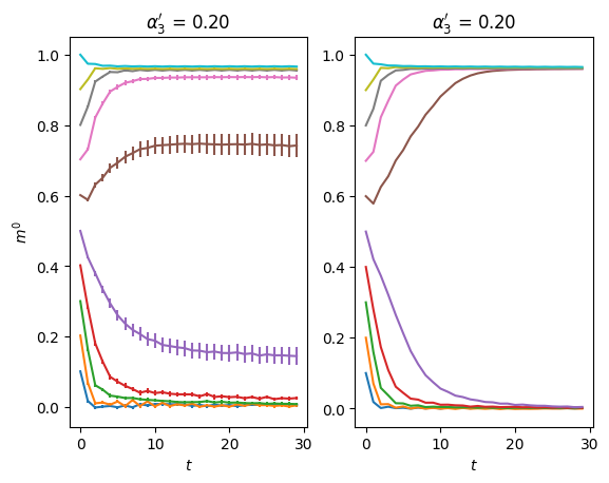} \\
	\small \hspace*{2em} (a) \hspace*{20em} (b) \\
	\includegraphics[width=0.45\linewidth,keepaspectratio]
	{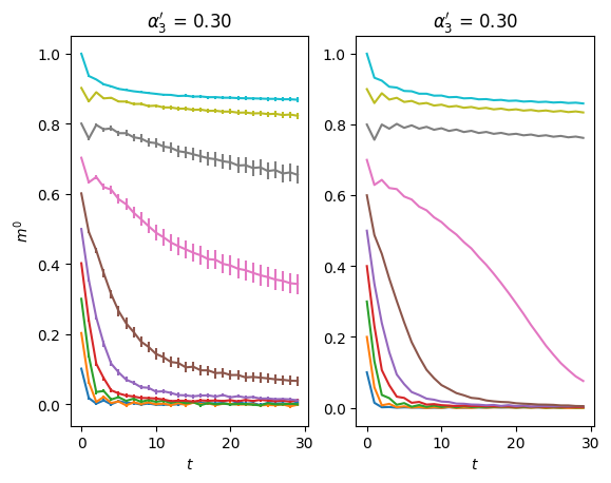} ~~~~
	\includegraphics[width=0.45\linewidth,keepaspectratio]
	{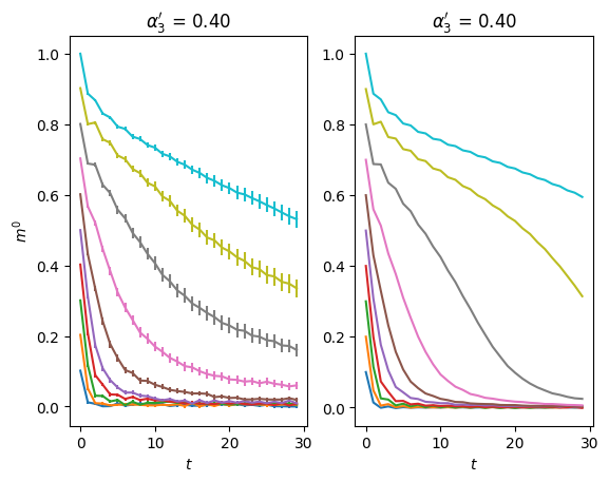} \\
	\small \hspace*{2em} (c) \hspace*{20em} (d) \\
	\caption{Recalling process of Krotov's dense associative memory with $F(x)=x^n$ and $n=3$. 
	Left: computer simulations, 100 trials, $N=512$. 
	Right: theory. 
	(a) $\alpha_3'=0.1$. 
	(b) $\alpha_3'=0.2$. 
	(c) $\alpha_3'=0.3$. 
	(d) $\alpha_3'=0.4$. }
	\label{fig:dynamics}
	\vskip2em
	\includegraphics[width=0.45\linewidth,keepaspectratio]
	{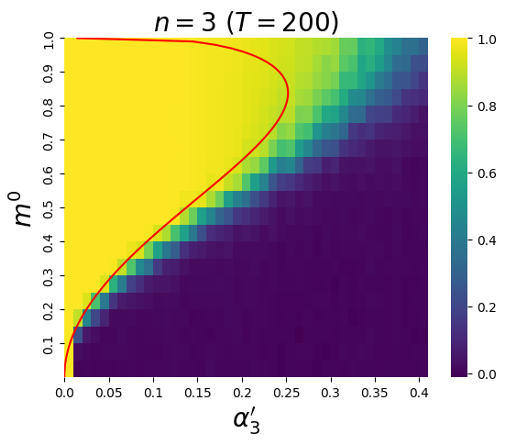} ~~~~
	\includegraphics[width=0.45\linewidth,keepaspectratio]
	{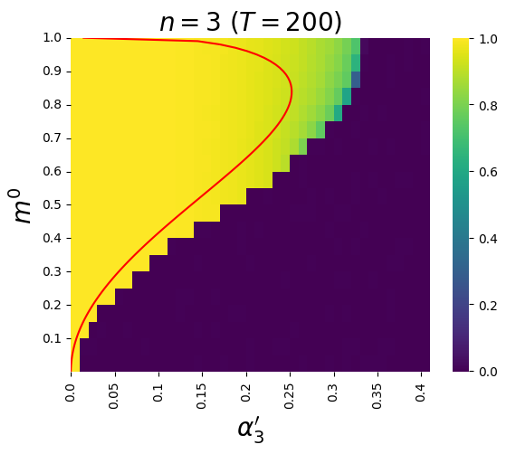} \\
	\caption{Attraction basin of Krotov's dense associative memory with $F(x)=x^n$ and $n=3$. 
	Left: computer simulations, 100 trials, $T=200$, $N=512$. 
	Right: theory, $T=200$. 
	Red line: attraction basin obtained by approximate dynamics}
	\label{fig:BoA}
\end{figure}

%--------------------------------------------------
\section{Discussion}
\par

%--------------------------------------------------
\subsection{Numerical Analysis and Computer Simulations}
\par
In this paper we considered Krotov's dense associative memory. Since the noise variance depends on $n$, we normalize the constant $\alpha_n$ by setting $\alpha_n'=(2n-3)!! \alpha_n$, where $\alpha_n'$ is referred to as the \textit{loading rate}. The \textit{storage capacity} $\alpha_{c,n}'$ is defined as the largest loading rate at which the overlap remains positive. 

\par
The result of Proposition \ref{proposition:gfa} can be numerically analyzed using the Monte Carlo method. Figure \ref{fig:dynamics} shows the numerical analysis for the case $n=3$, together with the results of computer simulations. In Fig. \ref{fig:dynamics} (a) -- (d), the graphs on the left display the simulation results for $N=512$ with $100$ trials. The vertical axis represents the overlap, while the horizontal axis represents the number of iteration steps. The graphs on the right in Fig. \ref{fig:dynamics}  (a) -- (d) correspond to the numerical analysis of the overlap based on Proposition \ref{proposition:gfa}. Although finite-size effects become significant near the basin of attraction, it can be confirmed that the theoretical values agree well with the simulation results even for relatively small-scale experiments. It can be theoretically confirms that, when retrieval is successful, convergence is attained within several tens of iterations. 

\par
Furthermore, Fig. \ref{fig:BoA} illustrates the basin of attraction for the case $n=3$. The vertical axis indicates the initial overlap, and the horizontal axis indicates the loading rate. The overlap after $200$ iterations is shown by color. The storage capacity is $\alpha_{c,3}' = 0.33$. The region in which the overlap remains finite and nonzero corresponds to the basin of attraction. The red solid line in Fig. \ref{fig:BoA} represents the boundary of the basin of attraction obtained from the approximate dynamics. These results demonstrate that the retarded self-interaction increases the memory capacity. It can be confirmed that the system retains a sufficiently large attraction basin of order $O(N^0)$, even for large $n$.

%--------------------------------------------------
\subsection{Connection to Related Analyses}
\par
From the exact solution obtained via the generating functional analysis, we obtain the following approximated result when self-coupling is neglected. 

\begin{corollary} % COROLLARY
	Neglecting the retarded self-interaction term as an approximation, i.e., setting $\Gamma=O$, we obtain 
	\begin{align}
		m^{(t+1)} = 
		\erf \biggl( \frac{(m^{(t)})^{n-1}}{\sqrt{(2n-3)!! ~ 2 \alpha_n}} \biggr), 
	\end{align}
	where $\erf(x) := \frac2{\sqrt{\pi}} \int_0^x e^{-z^2} dz$ 
	denotes the error function, and $m^{(0)}$ is the initial overlap. 
\end{corollary}

In this approximation, the equilibrium state of the dynamics can be simply obtained by setting $m^{(t)}=m$, and the resulting fixed-point equation corresponds to the equilibrium analysis by $n$-body Hopfield model. Although the coefficients differ, this is due to the fact that the energy function is not the same as that in Krotov's model. The storage capacity that obtained by the fixed-point equation of this approximated dynamics, i.e., $m = \erf ( m^{n-1}/(\sqrt{(2n-3)!! ~ 2 \alpha_n}) )$, gives that of the equilibrium analysis derived by the replica method. 

\par
We here consider the differences between Krotov's dense associative memory and the $n$-body Hopfield model independently proposed by Gardner and Abbott. The energy function of the $n$-body Hopfield model is defined by 
\begin{equation}
	H = - \frac1{\sqrt{2 n!} N^{n-1}} 
	\sum_{\mu=1}^M 
	\sum_{j_1 \ne j_2 \ne \cdots \ne j_n}^N 
	\xi_{j_1}^\mu 
	\xi_{j_2}^\mu \cdots 
	\xi_{j_n}^\mu ~
	h_{j_1} 
	h_{j_2} \cdots 
	h_{j_n} . 
\end{equation}
The values of the variables $j_1, \cdots, j_n$ are all distinct. This is the main difference from the Krotov's dense associative memory. Using the same way to Krotov's method, the corresponding update rule of the $n$-body Hopfield model is given as 
\begin{align}
	h_i^{(t+1)} = \sgn \biggl[
	\sum_{\mu=1}^M 
	\xi_i^\mu \frac{1}{N^{n-1}} 
	\sum_{j_1 \ne \cdots \ne j_{n-1} \ne i}^N
	\xi_{j_1}^\mu \cdots
	\xi_{j_{n-1}}^\mu ~ 
	h_{j_1}^{(t)} \cdots
	h_{j_{n-1}}^{(t)} 
	\biggr]. 
\end{align}
We have the exact result in the same way to obtain Proposition \ref{proposition:gfa}. Let $M=\alpha_n N^{n-1}$ again. Neglecting the retarded self-interaction term as an approximation, i.e., setting $\Gamma=O$, we obtain 
\begin{align}
	m^{(t)} = \erf \biggl( \frac{(m^{(t-1)})^{n-1}}{\sqrt{(n-1)! ~ 2 \alpha_n}} \biggr). 
	\label{eq:gardners-result}
\end{align}
The stationary equation, i.e., setting $m^{(t)}=m$, is equivalent to the result derived by Abbott \citep{Abbott1987}. The detail is available in Appendix \ref{appendix:Gardner}.

%--------------------------------------------------
\section{Conclusion}
\par
We performed an asymptotically exact analysis of the dynamical behaviour of dense associative memory using generating functional analysis (GFA) in the large-system limit. The analysis revealed the presence of a retarded self-coupling term, indicating that the next state of the system depends in a complex manner on all past states. We also confirmed that this property cannot be captured by a method based on the signal-to-noise analysis. In the traditional Hopfield model, i.e., $n=2$, it was found that the system exhibits a noise variance that depends intricately on non-recalled patterns. In contrast, for $n \ge 3$, the noise variance due to non-recalled patterns does not depend on the overlap with the recalled pattern. As a result, the phenomenon observed in the classical Hopfield model, namely, the increase in noise variance upon successful retrieval, is mitigated. Thus, it arises only from the retarded self-interaction. As a result, the recall process becomes simpler than the traditional Hopfield model. 
\par
Assuming the existence of a stationary state, we can also consider a macroscopic fixed-point equation from the GFA equations. Due to the presence of the self-coupling term, this result must differ from that of existing equilibrium analysis. This difference comes from the fact that, for models with $n \ge 3$, the system does not satisfy the detailed balance condition. 
\par
In this work, we provided an exact dynamical analysis of dense associative memory using the generating functional analysis, and verified the theoretical predictions with numerical experiments. Our results clarify how higher-order interactions, namely, $n \ge 3$, suppress the increasing of crosstalk noise due to the recalling pattern itself, thereby stabilizing recall dynamics and enhancing memory capacity. This contrasts with the classical Hopfield model, where self-retrieval inevitably introduces additional noise.
These findings offer a quantitative framework to evaluate the stability and storage capacity of associative memory models, which is useful for guiding model design. While our experiments were limited to relatively small system sizes and specific interaction orders, the analytical methodology is general and can be applied to a broader class of energy-based models. This approach can be extending to modern Hopfield networks, memory-augmented architectures, and other energy-based formulations will provide further insights into the design of robust and scalable memory systems.

%\subsubsection*{Author Contributions}
\subsubsection*{Acknowledgments}
\par
This work was partially supported by 
JSPS KAKENHI Grant Nos. 23H05492 (KM, JT) and 23K03841 (KM), 
and MEXT/JSPS KAKENHI Grant No. 22H05117 (YK). 

\bibliographystyle{iclr2026_conference.bst}
\bibliography{iclr2026_refs.bib}

% A number of width problems arise when LaTeX cannot properly hyphenate a line. 
% Please give LaTeX hyphenation hints using the \verb+\-+ command.

%--------------------------------------------------
\appendix
\newpage
\section*{Appendices}

%--------------------------------------------------
\section{Proof Sketch of Lemma \ref{lemma:Z}} 
\label{appendix:Z}
\par
We first calculate the expectation value of the noise term of all non-recalled patterns, which is the last part in (\ref{eq:noise-in-Z}). Using the Taylor expansion, we obtain 
\begin{align}
	&
	\mathbb{E}_{\vec{\xi}^2, \cdots, \vec{\xi}^M}	
	\exp \biggl[ - i \sum_{t=0}^{T-1} \sum_{i=1}^N 
		\sum_{\mu =2}^M 
		\hat{u}_i^{(t)} \xi_i^\mu ~ 
		n \biggl( \frac1N \sum_{j \ne i}^N \xi_j^\mu h_j^{(t)} \biggr)^{n-1}
	\biggr]
	\\
	=& 
	\prod_{\mu=2}^M \mathbb{E}_{\vec{\xi}^\mu}	
	\biggl\{ 1 
		+ \frac12 \biggl(- i \sum_{t=0}^{T-1} \sum_{i=1}^N 
		\hat{u}_i^{(t)} \xi_i^\mu ~ 
		n \biggl( \frac1N \sum_{j \ne i}^N \xi_j^\mu h_j^{(t)} \biggr)^{n-1} \biggr)^2
		+ O\biggl( \frac{n^3}{N^{3(n-1)}} \biggr)
	\biggr\}
	\\
	=&
	\exp \biggl[ 
	- 
	\frac{n^2(M-1)}{2N^{2(n-1)}} \biggl( 
		\sum_{t=0}^{T-1} \sum_{t'=0}^{T-1} \sum_{i=1}^{N } 
		\hat{u}_i^{(t)} \hat{u}_i^{(t')} \mathcal{N}_1 + \!
		\sum_{t=0}^{T-1} \sum_{t'=0}^{T-1} \sum_{i=1}^{N } \sum_{i' \ne i}^{N} 
		\hat{u}_i^{(t)} \hat{u}_{i'}^{(t')} \mathcal{N}_2 \!
	\biggr)
	\!+\! O\biggl(\! \frac{n^3 M}{N^{3(n-1)}} \!\biggr)
	\biggr], 
	\label{eq:appendix-N}
\end{align}
where 
\begin{align}
	\mathcal{N}_1
	=& 
	\mathbb{E}_{\vec{\xi}} \biggl[
	\sum_{j_1  \ne i }^N \cdots\!\!\! \sum_{j_{n-1}  \ne i }^N 
	\sum_{j_1' \ne i'}^N \cdots\!\!\! \sum_{j_{n-1}' \ne i'}^N 
	\xi_{j_1 } \cdots \xi_{j_{n-1} } 
	\xi_{j_1'} \cdots \xi_{j_{n-1}'} 
	h_{j_1 }^{(t )} \cdots h_{j_{n-1} }^{(t )} 
	h_{j_1'}^{(t')} \cdots h_{j_{n-1}'}^{(t')} 
	\biggr], 
	\notag\\
	\mathcal{N}_2
	=& 
	\mathbb{E}_{\vec{\xi}}\biggl[
	\xi_i \xi_{i'} 
	\sum_{j_1  \ne i }^N \cdots\!\!\! \sum_{j_{n-1}  \ne i }^N 
	\sum_{j_1' \ne i'}^N \cdots\!\!\! \sum_{j_{n-1}' \ne i'}^N 
	\xi_{j_1 } \cdots \xi_{j_{n-1} } 
	\xi_{j_1'} \cdots \xi_{j_{n-1}'} 
	h_{j_1 }^{(t )} \cdots h_{j_{n-1} }^{(t )} 
	h_{j_1'}^{(t')} \cdots h_{j_{n-1}'}^{(t')} 
	\biggr]. \notag
\end{align}
\par
Since $\vec{\xi}^2, \cdots, \vec{\xi}^M$ are independent, we can drop the index $\mu$. 
It should be noted that any term in $\mathcal{N}_1$ and $\mathcal{N}_2$ that contains an odd number of identical index from the same pattern has zero expectation, because all $\xi_1 ,\cdots, \xi_N$ are independent and have zero mean. 
% (*2) is expected to be $O(N^n)$. If so, $(*1) \!\!=\!\! \{e^{-\frac1{2N^{2(n-1)}} (*2)}\}^{M-1} \!$ becomes $O(e^N)$, when $M=O(N^{n-1})$. (*4) is $O(N^{n-1})$. So, at least the second term becomes $O(N^n)$ due to the summation of $i$. So, (*2) become $O(N^n)$. Then, (*1) is $O(e^N)$ if $M=O(N^{n-1})$. 
\par
We calculate $\mathcal{N}_1$. The leading term in $\mathcal{N}_1$ can be obtained by calculating the summations in the case where the $2(n-1)$ variables are grouped into pairs, each pair taking the same value. We have to do this for all possible partitions. We must distinguish three types of pairings: (i) between two primed variables, (ii) between two unprimed variables, and (iii) between a primed and an unprimed variable. Note that depending on the type of pair, the time parameter differs. Therefore, the leading term can be obtained by counting the number of ways to partition the $2(n-1)$ indices, i.e., $j_1, \cdots, j_{n-1}, j_1', \cdots, j_{n-1}'$, into $n-1$ pairs in which indices take the same value while each different pairs takes different values. 
\par
We here consider two sets of indices: the set of unprimed indices $\mathcal{J} = \{ j_1, \cdots, j_{\ell} \}$ and the set of primed indices $\mathcal{J}' = \{ j_1' , \cdots, j_{\ell}' \}$. First, we consider the number of ways to divide $2 \ell$ indices, including $\ell$ unprimed indices and $\ell$ primed indices, into $\ell$ pairs. Let $A(\ell,k)$ be the number of ways to have exactly $k$ unprimed-primed pairs in $\ell$ total pairs, which is given by 
\begin{equation}
	A(\ell,k)
	=
	\left(
	\begin{array}{c}
		\ell \\
		k \\
	\end{array}
	\right)^2 \; k! \; B(\ell-k)^2, 
\end{equation}
where $B(m)$ is the number of ways where $m/2$ unprimed-unprimed pairs and $m/2$ primed-primed pairs are made using $2m$ indices, consisting of $m$ unprimed indices and $m$ primed indices: 
\begin{align}
	B(m) 
	= 
	\vec{1}_{m\mbox{:even}} \; 
	\frac{1}{(m/2)!}
	\left(
	\begin{array}{c}
		m \\
		2 \\
	\end{array}
	\right)
	\left(
	\begin{array}{c}
		m-2 \\
		2 \\
	\end{array}
	\right)
	\cdots
	\left(
	\begin{array}{c}
		2 \\
		2 \\
	\end{array}
	\right) 
	= 
	\vec{1}_{m\mbox{:even}} \; (m-1)!!. 
\end{align}
Note that $\sum_{k=0}^{\ell} A(\ell,k) = B(2\ell)$ holds. 
\par
Using the quantity $A(\ell,k)$ and the identity $(h_{i}^{(t)})^2=1$, we obtain 
\begin{align}
	\mathcal{N}_1
	=& 
	\mathbb{E}_{\vec{\xi}} \biggl[
	\sum_{j_1, \cdots, j_{n-1} \ne i}^N 
	\sum_{j_1', \cdots, j_{n-1}' \ne i}^N 
	\xi_{j_1} \cdots \xi_{j_{n-1}} 
	\xi_{j_1'} \cdots \xi_{j_{n-1}'} 
	h_{j_1}^{(t)} \cdots h_{j_{n-1}}^{(t)} 
	h_{j_1'}^{(t')} \cdots h_{j_{n-1}'}^{(t')} 
	\biggr]
	\\
	=& 
	N^{n-1} ~
	\underbrace{
	\sum_{k=0}^{n-1}
	A(n-1,k) 
	\biggl(
		\frac1N \sum_{j=1}^N h_j^{(t)} h_j^{(t')}
	\biggr)^k
	}_{=O(N^0)} ~ 
	+ O(N^{n-3}). 
	\label{eq:appendix-N1}
\end{align}
\par
Next, we calculate $\mathcal{N}_2$. For notational simplicity,let $(\; \cdot \; |_{j_1=i'}) (\cdots)$ be an operator to substitute $j_1=i'$ into $(\cdots)$, and let $(\sum_{j_1 \ne i,\ne i'}\cdot \;) (\cdots)$ be an operator for summing  $(\cdots)$ over $j_1 \ne i,\ne i'$. It should be noted that each of $j_1, \cdots, j_{n-1}$ can take the value $i'$, and conversely, each of $j_1', \cdots, j_{n-1}'$ can take the value $i$. For all $i \in \{1, \cdots, N\}$ and $i' \in \{1, \cdots, N\} \backslash \{i\}$, we have 
\begin{align}
	\mathcal{N}_2
	%=& 
	%\mathbb{E}_{\vec{\xi}}\biggl[
	%\xi_i \xi_{i'} 
	%\sum_{j_1 \ne i}^N \cdots \sum_{j_{n-1} \ne i}^N ~ 
	%\sum_{j_1' \ne i'}^N \cdots \sum_{j_{n-1}' \ne i'}^N ~ 
	%\xi_{j_1} \cdots \xi_{j_{n-1}} 
	%\xi_{j_1'} \cdots \xi_{j_{n-1}'} 
	%h_{j_1}^{(t)} \cdots h_{j_{n-1}}^{(t)} 
	%h_{j_1'}^{(t')} \cdots h_{j_{n-1}'}^{(t')} 
	%\biggr]
	%\notag\\
	=& 
	\mathrm{E}_{\bm{\xi}}\biggl[
	\xi_i \xi_{i'} 
	\biggl(\;\cdot\;\biggl.\biggr|_{j_1=i'} + \sum_{j_1 \ne i,\ne i'}\cdot\;\biggr) \cdots 
	\biggl(\;\cdot\;\biggl.\biggr|_{j_{n-1}=i'} + \sum_{j_{n-1} \ne i,\ne i'}\cdot\;\biggr) 
	\nonumber\\
	&
	~~~~~~~~~~~~~~
	\biggl(\;\cdot\;\biggl.\biggr|_{j_1'=i} + \sum_{j_1' \ne i',\ne i}\cdot\;\biggr) \cdots 
	\biggl(\;\cdot\;\biggl.\biggr|_{j_{n-1}'=i} ~+ \sum_{j_{n-1}' \ne i',\ne i}\cdot\;\biggr) 
	\nonumber\\
	&
	~~~~~~~~~~~~~~
	\xi_{j_1} \cdots \xi_{j_{n-1}} 
	\xi_{j_1'} \cdots \xi_{j_{n-1}'} 
	h_{j_1}^{(t)} \cdots h_{j_{n-1}}^{(t)} 
	h_{j_1'}^{(t')} \cdots h_{j_{n-1}'}^{(t')} 
	\biggr]
	\\
	=& 
	\left(
	\begin{array}{c}
		n-1 \\
		1 \\
	\end{array}
	\right)
	\left(
	\begin{array}{c}
		n-1 \\
		1 \\
	\end{array}
	\right)
	\mathrm{E}_{\bm{\xi}}\biggl[
	\xi_i \xi_{i'}
	\nonumber\\
	&
	~~~~~~~~~~
	\biggl(\sum_{j_1 \ne i,\ne i'}\cdot\;\biggr) \cdots 
	\biggl(\sum_{j_{n-2} \ne i,\ne i'}\cdot\;\biggr) 
	\biggl(\;\cdot\;\biggl.\biggr|_{j_{n-1}=i'} \biggr) 
	\nonumber\\
	&
	~~~~~~~~~~
	\biggl(\sum_{j_1' \ne i',\ne i}\cdot\;\biggr) \cdots 
	\biggl(\sum_{j_{n-2}' \ne i',\ne i}\cdot\;\biggr) 
	\biggl(\;\cdot\;\biggl.\biggr|_{j_{n-1}'=i} ~\biggr) 
	\nonumber\\
	&
	~~~~~~~~~~
	\xi_{j_1} \cdots \xi_{j_{n-1}} 
	\xi_{j_1'} \cdots \xi_{j_{n-1}'} 
	h_{j_1}^{(t)} \cdots h_{j_{n-1}}^{(t)} 
	h_{j_1'}^{(t')} \cdots h_{j_{n-1}'}^{(t')} 
	\biggr]
	+O(N^{n-4})
	\\
	=& 
	(n-1)^2 ~~ 
	h_{i'}^{(t)} h_{i}^{(t')} ~~  
	\mathrm{E}_{\bm{\xi}}\biggl[
	\sum_{j_1, \cdots, j_{n-2} \ne i,\ne i'} ~ 
	\sum_{j_1', \cdots, j_{n-2}' \ne i',\ne i} 
	\nonumber\\
	&
	~~~~~~~~~~
	\xi_{j_1} \cdots \xi_{j_{n-2}} 
	\xi_{j_1'} \cdots \xi_{j_{n-2}'} 
	h_{j_1}^{(t)} \cdots h_{j_{n-2}}^{(t)} 
	h_{j_1'}^{(t')} \cdots h_{j_{n-2}'}^{(t')} 
	\biggr]
	+O(N^{n-4})
	\\
	=&
	(n-1)^2 ~~ h_{i'}^{(t)} h_{i}^{(t')} ~~ 
	N^{n-2} ~
	\underbrace{
	\sum_{k=0}^{n-2}
	A(n-2,k) 
	\biggl(
		\frac1N \sum_{j=1}^N h_j^{(t)} h_j^{(t')}
	\biggr)^k
	}_{=O(N^0)} ~ 
	+ O(N^{n-4}). 
	\label{eq:appendix-N2}
\end{align}
\par
Substituting (\ref{eq:appendix-N1}) and (\ref{eq:appendix-N2}) into (\ref{eq:appendix-N}), we obtain the expectation value of the noise term of all non-recalled patterns as follows: 
\begin{align}
	&
	\mathbb{E}_{\vec{\xi}^2, \cdots, \vec{\xi}^M}	
	\exp \biggl[ - i \sum_{t=0}^{T-1} \sum_{i=1}^N 
		\sum_{\mu =2}^M 
		\hat{u}_i^{(t)} \xi_i^\mu ~ 
		n \biggl( \frac1N \sum_{j \ne i}^N \xi_j^\mu h_j^{(t)} \biggr)^{n-1}
	\biggr]
	\\
	=& 
	\exp \biggl[
	-\frac12 \cdot \frac{n^2 M}{N^{n-2}} ~
	\sum_{t=0}^{T-1} \sum_{t'=0}^{T-1} \biggl\{ 
	\nonumber\\
	& 
	(n-1)^2 
	\biggl(\frac1N \sum_{i =1}^N h_{i }^{(t')} \hat{u}_{i }^{(t )} \biggr)
	\biggl(\frac1N \sum_{i'=1}^N h_{i'}^{(t )} \hat{u}_{i'}^{(t')} \biggr)
	\sum_{k=0}^{n-2}
	A(n-2,k) 
	\biggl(
		\frac1N \sum_{j=1}^N h_j^{(t)} h_j^{(t')}
	\biggr)^k
	\nonumber\\
	& 
	+ 
	\biggl( \frac1N \sum_{i=1}^N \hat{u}_i^{(t)} \hat{u}_{i}^{(t')} \biggr) 
	\sum_{k=0}^{n-1}
	A(n-1,k) 
	\biggl(
		\frac1N \sum_{j=1}^N h_j^{(t)} h_j^{(t')}
	\biggr)^k 
	+ O(N^{-1}) \biggr\} 
	\biggr]. 
	\label{eq:app-noise}
\end{align}
\par
The signal term that includes the recalling pattern $\vec{\xi}^1$ can be rearranged as 
\begin{align}
	&
	\mathbb{E}_{\vec{\xi}^1} 
	\exp \biggl[ - i \sum_{t=0}^{T-1} \sum_{i=1}^N 
	\hat{u}_i^{(t)} \xi_i^1~ 
	n \biggl( \frac1N \sum_{j \ne i}^N \xi_j^1 h_j^{(t)} \biggr)^{n-1}
	\biggr] \notag\\
	=& 
	\mathbb{E}_{\vec{\xi}^1} 
	\exp \biggl[ - i N \sum_{t=0}^{T-1} 
	\biggl(\frac1N \sum_{i=1}^N \hat{u}_i^{(t)} \xi_i^1 \biggr)~ 
	n \biggl( \frac1N \sum_{j=1}^N \xi_j^1 h_j^{(t)} + O(N^{-1}) \biggr)^{n-1} 
	\biggr]. 
	\label{eq:app-signal}
\end{align}
Since the expectation over the non-recalled patterns has already been taken, and only the recalling pattern remains in the expression. The signal term of (\ref{eq:app-signal}) is $e^{O(N)}$. On the other hand, the noise term of (\ref{eq:app-noise}) is $e^{O(M/N^{n-2})}$. For non-trivial analysis, the signal term and the noise term must be of the same order, namely, the number of the memory patterns $M$ must be $O(N^{n-1})$. 
\par
We introduce the parameters of (\ref{eq:macroscopic-parameters}) into (\ref{eq:app-noise}) by the Dirac delta function. Using the Fourier integral form of the Dirac delta function, we arrive at the generating functional of (\ref{eq:bar-Z-in-lemma}).

%--------------------------------------------------
\section{Proof Sketch of Proposition \ref{proposition:gfa}}
\label{appendix:gfa}
\par
It should be noted that the normalization relation $Z[\vec{0}]=1$ plays an important role in the elimination of spurious solutions to the saddle-point equations. The terms in the averaged generating functional can be split into three related parts. The first one is a signal part. The second one is a static noise part due to the random variables within the model. The last one is retarded self-interaction due to the influence of the state at the previous stage, which may be able to affect the present state. The GFA allows us to treat the last part. After the analysis, it turns out that the system can be described in terms of the following three quantities: 
\begin{align}
	& 
	m^{(t)} = 
	\mathbb{E}_{\vec{\xi}^1, \cdots, \vec{\xi}^M} 
	\biggl[ \biggl\langle
	\frac1N \sum_{i=1}^N \xi_i^1 h_i^{(t)}
	\biggr\rangle \biggr], \\
	&
	C^{(t,t')} = 
	\mathbb{E}_{\vec{\xi}^1, \cdots, \vec{\xi}^M} 
	\biggl[ \biggl\langle
	\frac1N \sum_{i=1}^N h_i^{(t)} h_i^{(t')} 
	\biggr\rangle \biggr], \\
	&
	G^{(t,t')} = 
	\mathbb{E}_{\vec{\xi}^1, \cdots, \vec{\xi}^M} 
	\biggl[ \biggl\langle
	\frac1N \sum_{i=1}^N \frac
	{\partial h_i^{(t)}}
	{\partial \theta_i^{(t')}} 
	\biggr\rangle \biggr], 
\end{align}
where these are referred to as the \textit{overlap}, the \textit{correlation function}, and the \textit{response function}, respectively. One can deduce the meaning of macroscopic parameters by differentiating the averaged generating functional with respect to the external field $\theta_i^{(t)}$ and generating functions $\psi_i^{(t)}$. The averaged generating functional $\bar{Z}[\vec{\psi}]$ is dominated by a saddle-point for $N\to\infty$. Using the normalization identity $\bar{Z}[\vec{0}]=\mathbb{E}_{\vec{\xi}^1, \cdots, \vec{\xi}^M} \langle 1 \rangle=1$, one can have derivatives of the averaged generating functional: 
\begin{align}
	& 
	\lim_{\vec{\psi}\to\vec{0}} 
	\frac{\partial \bar{Z}[\vec{\psi}]}
	{\partial \psi_i^{(t)}} 
	= 
	-\rmi \langle h^{(t)} \rangle_i, 
	\notag\\
	& 
	\lim_{\vec{\psi}\to\vec{0}} 
	\frac{\partial^2 \bar{Z}[\vec{\psi}]}
	{\partial \psi_i^{(t)} \partial \psi_{i'}^{(t')}} 
	= 
	-\delta_{i,i'} 
	\langle h^{(t)} h^{(t')}\rangle_i 
	- (1-\delta_{i,i'}) 
	\langle h^{(t)}\rangle_i 
	\langle h^{(t')}\rangle_{i'}, 
	\notag\\
	& 
	\lim_{\vec{\psi}\to\vec{0}} 
	\frac{\partial^2 \bar{Z}[\vec{\psi}]}
	{\partial \psi_i^{(t)} \partial \theta_{i'}^{(t')}}
	= 
	-\delta_{i,i'} 
	\langle h^{(t)} \hat{u}^{(t')}\rangle_i 
	- (1-\delta_{i,i'}) 
	\langle h^{(t)}\rangle_i 
	\langle \hat{u}^{(s')}\rangle_{i'}
	\label{eq:derivatives2}
	\\
	&
	\lim_{\vec{\psi}\to\vec{0}} 
	\frac{\partial \bar{Z}[\vec{\psi}]}
	{\partial \theta_i^{(t)}} 
	= -\rmi\langle \hat{u}^{(t)}\rangle_i
	= 0
	\notag\\
	& 
	\lim_{\vec{\psi}\to\vec{0}} 
	\frac{\partial^2 \bar{Z}[\vec{\psi}]}
	{\partial \theta_i^{(t)} \partial \theta_{i'}^{(t')}} 
	= -\delta_{i,i'}\langle \hat{u}^{(t)}\hat{u}^{(t')}\rangle_i
	= 0, 
	\notag
\end{align}
where $\langle \; \rangle_i$ denotes the average that is defined by 
\begin{align}
	& \langle f(\vec{h},\vec{u},\hat{\vec{u}}) \rangle_i := 
	\frac {\displaystyle \sum_{\vec{h}} \int d\vec{u} d\hat{\vec{u}} ~
	w_i(\vec{h},\vec{u},\hat{\vec{u}})
	f(\vec{h},\vec{u},\hat{\vec{u}})}
	{\displaystyle \sum_{\vec{h}} \int d\vec{u} d\hat{\vec{u}} ~
	w_i(\vec{h},\vec{u},\hat{\vec{u}})}
\end{align}
with 
\begin{align}
	w_i(\vec{h},\vec{u},\hat{\vec{u}})
	=&
	p[h^{(0)}]
	\biggl( \prod_{t=0}^{T-1} \delta[ h^{(t+1)} ; \sgn(u^{(t)}) ] \biggr) 
	\notag\\
	& 
	- i \sum_{t=0}^{T-1} \sum_{t'=0}^{T-1} \{ 
	\hat{q}^{(s,s')} h^{(s)} \tilde{b}^{(s')} 
	+ \hat{Q}^{(s,s')} \hat{u}^{(s)} \hat{u}^{(s')} 
	+ \hat{K}^{(s,s')} h^{(s)} \hat{u}^{(s')} \} 
	\notag\\
	&
	+ i \sum_{t=0}^{T-1} \hat{u}^{(t)} 
	\{ u^{(t)} - \hat{k}^{(t)} \xi_i - \theta_i^{(t)} \} 
	- i \sum_{t=0}^{T-1} h^{(s)} \hat{m}^{(t)} \biggr] 
	\biggl. \biggr|_{\mathrm{saddle}}. 
\end{align}
The average $\langle (\cdots) \rangle_i$ is referred to as a {\it single-unit measure}. Here, evaluation $f|_{\mathrm{saddle}}$ denotes an evaluation of function $f$ at the dominating saddle-point. Substituting (\ref{eq:derivatives2}) into (\ref{eq:derivatives.1}) -- (\ref{eq:derivatives.3}), we then have 
\begin{align}
	& 
	\mathbb{E}_{\vec{\xi}^1, \cdots, \vec{\xi}^M} 
	\langle h_i^{(t)} \rangle
	= 
	\langle h^{(t)} \rangle_i, 
	\notag\\
	&
	\mathbb{E}_{\vec{\xi}^1, \cdots, \vec{\xi}^M} 
	\langle h_i^{(t)} h_{i'}^{(t')} \rangle
	= 
	\delta_{i,i'} \langle h^{(t)} h^{(t')} \rangle_i 
	+ (1-\delta_{i,i'}) \langle h^{(t)} \rangle_i \langle h^{(t')} \rangle_{i'}, 
	\label{eq:derivatives3}
	\\
	& 
	\mathbb{E}_{\vec{\xi}^1, \cdots, \vec{\xi}^M} 
	\langle 
	\frac
	{\partial h_i^{(t)}}
	{\partial \theta_{i'}^{(t')}}
	\rangle 
	= 
	- i \delta_{i,i'} \langle h^{(t)} \hat{u}^{(t')}\rangle_i. 
	\notag
\end{align}
\par
In the large-system limit, the averaged generating functional will be evaluated by the dominating saddle-points of the exponent $\Phi+\Psi+\Omega$. We can now derive the saddle-point equations by differentiation with respect to the integral variables $m^{(t)}$, $\hat{m}^{(t)}$, $k^{(t)}$, $\hat{k}^{(t)}$, $q^{(t,t')}$, $\hat{q}^{(t,t')}$, $Q^{(t,t')}$, $\hat{Q}^{(t,t')}$, $K^{(t,t')}$, and $\hat{K}^{(t,t')}$. The saddle-point equations will involve the overlap $m^{(t)}$, the correlation $C^{(t,t')}$ and the response function $G^{(s,s')}$. It should be noted that causality, i.e., 
\begin{align}
  \frac
  {\partial \langle h^{(t)} \rangle}
  {\partial \theta^{(t')}}
  =0, 
\end{align}
should hold for $t \le t'$. Therefore $G^{(t,t')}=0$ for $t \le t'$. Using causality and the identities (\ref{eq:derivatives2}) and (\ref{eq:derivatives3}), the straightforward differentiation of $\Phi+\Psi+\Omega$ with respect to the integral variables leads us to the following saddle-point equations: 
\begin{align}
	&
	m^{(t)} 
	= \frac1N \sum_{i=1}^N \xi_i^1 \overline{\langle h_i^{(t)} \rangle} 
	= \llangle \xi h^{(t)} \rrangle, ~~~~ 
	\hat{m}^{(t)}=0, ~~~~ 
	k^{(t)} =0, ~~~~ 
	\hat{k}^{(t)} = n(m^{(t)})^{n-1}, 
	\\
	&
	q^{(t,t')} 
	= C^{(t,t')}
	= \frac1N \sum_{i=1}^N \overline{\langle h_i^{(t)} h_i^{(t')} \rangle} 
	= \llangle h^{(t)} h^{(t')} \rrangle, ~~~~ 
	\hat{q}^{(t,t')} = 0, 
	\\
	& 
	Q^{(t,t')}=0, ~~~~ 
	\hat{Q}^{(t,t')} = - i \frac12 R^{(t,t')}, \\
	& 
	K^{(t,t')}= i G^{(t,t')} = \vec{1}_{t>t'}  
	\frac
	{\partial \llangle h^{(t)} \rrangle}
	{\partial \theta^{(t)}}, ~~~~ 
	\hat{K}^{(t,t')} = D^{(t,t')} G^{(t',t)}, 
\end{align}
where 
\begin{align}
	& R^{(t,t')} = n^2 \alpha_n \sum_{k=0}^{n-1} A(n-1,k) (C^{(t,t')})^k, \\
	& D^{(t,t')} = n^2 (n-1)^2 \alpha_n \sum_{k=0}^{n-2} A(n-2,k) (C^{(t,t')})^k. 
\end{align}
Substituting the solutions of the saddle-point equation into the single-unit measure, we obtain the effective path measure. We then arrive at Proposition \ref{proposition:gfa}.

%--------------------------------------------------
\section{Proof Sketch of Gardner's model}
\label{appendix:Gardner}
\par
Using a similar way to Lemma \ref{lemma:Z}, the expectation value of the noise term of all non-recalled patterns for the $n$-body Hopfield model is given by 
\begin{align}
	&
	\mathrm{E}_{\vec{\xi}^2, \cdots, \vec{\xi}^M}
	\exp \biggl[ - i \sum_{t=0}^{T-1} \sum_{i=1}^N 
	\sum_{\mu =2}^M 
	\hat{u}_i^{(t)} \xi_i^\mu 
	\frac1{N^{n-1}} 
	\sum_{j_1 \ne \cdots \ne j_{n-1} \ne i}^N 
	\xi_{j_1}^\mu \cdots
	\xi_{j_{n-1}}^\mu ~ 
	h_{j_1}^{(t)} \cdots
	h_{j_{n-1}}^{(t)} 
	\biggr] \notag\\
	=& 
	\exp \biggl[
	-\frac12 \frac{n^2 M}{N^{n-2}} 
	\sum_{t=0}^{T-1} \sum_{t'=0}^{T-1} \biggl\{
	(n-1)^2 
	\biggl( \frac1N \sum_{i =1}^N h_{i }^{(t')} \hat{u}_{i }^{(t )} \biggr) 
	\biggl( \frac1N \sum_{i'=1}^N h_{i'}^{(t )} \hat{u}_{i'}^{(t')} \biggr) 
	A(n-2,n-2) 
	\notag\\
	&\times 
	\biggl( \frac1N \sum_{j=1}^N h_j^{(t)} h_j^{(t')} \biggr)^{n-2}
	%\notag\\
	%&
	\!\!\!\!\!\!
	+ 
	\biggl( \frac1N \sum_{i=1}^N \hat{u}_i^{(t)} \hat{u}_{i}^{(t')} \biggr) 
	A(n-1,n-1) 
	\biggl( \frac1N \sum_{j=1}^N h_j^{(t)} h_j^{(t')}\biggr)^{n-1}
	\!\!\!\!\!\! + O(N^{-1}) \biggr\} 
	\biggr], 
\end{align}
where 
\begin{align}
	A(\ell,\ell)
	=
	\biggl(
	\!\!
	\begin{array}{c}
		\ell \\
		\ell \\
	\end{array}
	\!\!
	\biggr)^2 \; \ell! \; B(\ell-\ell)^2 = \ell!. 
\end{align}
Applying the same calculation, we arrive at (\ref{eq:gardners-result}).

\end{document}